\documentclass[12pt]{article}\usepackage[]{graphicx}\usepackage[]{color}
\makeatletter
\def\maxwidth{ %
  \ifdim\Gin@nat@width>\linewidth
    \linewidth
  \else
    \Gin@nat@width
  \fi
}
\makeatother

\definecolor{fgcolor}{rgb}{0.345, 0.345, 0.345}

\usepackage{framed}
\makeatletter
 {\par\unskip\endMakeFramed%
 \at@end@of@kframe}
\makeatother

\definecolor{shadecolor}{rgb}{.97, .97, .97}
\definecolor{messagecolor}{rgb}{0, 0, 0}
\definecolor{warningcolor}{rgb}{1, 0, 1}
\definecolor{errorcolor}{rgb}{1, 0, 0}
\newenvironment{knitrout}{}{} % an empty environment to be redefined in TeX

\usepackage{alltt}
\usepackage{graphicx}
\usepackage{color}
\usepackage{xcolor}
\usepackage{alltt}
\usepackage{pifont}
\usepackage{extarrows}% http://ctan.org/pkg/extarrows
\usepackage[toc,page]{appendix}
\usepackage{titling} %for the abstract on title page
\usepackage{authblk} %for author affiliations
\usepackage[margin=1in]{geometry} 
\usepackage{amsmath,amsthm,amssymb}
\usepackage{longtable}
\usepackage{array}
\usepackage{wrapfig}
\usepackage{colortbl}
\usepackage{pdflscape}
\usepackage{tabu}
\usepackage[normalem]{ulem}
\usepackage{makecell}
\usepackage{cite} %to cite multiple references
\usepackage{subfig}
\newcommand*{\mydprime}{^{\prime\prime}\mkern-1.2mu} %double prime
\providecommand{\keywords}[1]{\textbf{\textit{Keywords:}} #1} % define a command for keywords in abstract
\usepackage{changepage}
\usepackage{multirow}
\usepackage{bm}
\usepackage{float}
\usepackage{rotating}%for \sidewaystable
\definecolor{armygreen}{rgb}{0.5, 1, 0.13}

\usepackage[authoryear]{natbib}
\usepackage{float} %for fixing figure position to 'H'
\usepackage{enumitem}

\DeclareMathOperator{\Var}{Var} % Varianz
\DeclareMathOperator{\E}{\mathsf{E}} % Erwartungswert
\DeclareMathOperator{\SD}{SD} % standard deviation

\title{\bf Quantification of empirical determinacy: the impact of likelihood weighting on posterior location and spread in Bayesian meta-analysis estimated with JAGS and INLA}
 
\author[1]{Sona Hunanyan}
\author[2]{H\aa vard Rue}
\author[3]{Martyn Plummer}
\author[1]{Ma{\l}gorzata Roos}
\affil[1]{\small Department of Biostatistics at the Epidemiology, Biostatistics and Prevention Institute\\
University of Zurich, Hirschengraben 84, CH-8001 Zurich, Switzerland}
\affil[2]{\small CEMSE Division, King Abdullah University of Science and Technology, Saudi Arabia}
\affil[3]{Department of Statistics, University of Warwick, United Kingdom}

\date{\today}
 
\usepackage{hyperref}
\usepackage{mathtools}
\mathtoolsset{showonlyrefs} %number only those equations which are referred to

\IfFileExists{upquote.sty}{\usepackage{upquote}}{}
\IfFileExists{upquote.sty}{\usepackage{upquote}}{}
\begin{document}

\begin{titlingpage}
\maketitle
\begin{abstract}
The popular Bayesian meta-analysis expressed by Bayesian normal-normal hierarchical model (NNHM) synthesizes knowledge from several studies and is highly relevant in practice.
Moreover, NNHM is the simplest Bayesian hierarchical model (BHM), which illustrates problems typical in more complex BHMs.
Until now, it has been unclear to what extent the data determines the marginal posterior distributions of the parameters in NNHM.
To address this issue we computed the second derivative of the Bhattacharyya coefficient with respect to the weighted likelihood, defined the total empirical determinacy (TED), the proportion of the empirical determinacy of location to TED (pEDL), and the proportion of the empirical determinacy of spread to TED (pEDS).
We implemented this method in the R package \texttt{ed4bhm} and considered two case studies and one simulation study.
We quantified TED, pEDL and pEDS under different modeling conditions such as model parametrization, the primary outcome, and the prior.
This clarified to what extent the location and spread of the marginal posterior distributions of the parameters are determined by the data.
Although these investigations focused on Bayesian NNHM, the method proposed is applicable more generally to complex BHMs.

\vspace{5mm}

{\keywords Empirical determinacy, likelihood weighting, Bayesian meta-analysis, Bayesian hierarchical models, identification}
\end{abstract}
\end{titlingpage}

\section{Introduction}

Bayesian meta-analysis, which synthesizes the evidence from several already published studies, is an indispensable tool for evidence-based medicine.
Usually, Bayesian meta-analysis is based on a Bayesian normal-normal hierarchical model (NNHM) \citep{friede_roever_wandel_neuenschwander_OrphanDiseases_2017, friede_roever_wandel_neuenschwander_TwoStudies_2017, roever_2018, bender_friede_koch_kuss_schlattmann_schwarzer_skipka_2018}.
This is the simplest Bayesian hierarchical model (BHM), but presents problems typical in more complex BHMs \citep{vehtari_gelman_simpson_carpenter_buerkner_2021}. 

If different values of the parameters in a likelihood map to a single data model, then these values are indeterminate, or in other words non-identified, and are not estimable from the likelihood alone \citep{poirier_98, gustafson_15, lewbel_2019}.
The Bayesian approach is known to mitigate this problem, because a likelihood combined with the proper priors for the parameters that are not likelihood identifiable still yields posterior estimates \citep{kadane_1975, gelfand_sahu_99, eberly_carlin_00, gustafson_15}.
This raises a legitimate question: to what extent are Bayesian posterior parameter estimates determined by the data?

In classical statistical models, non-identified parameters in the likelihood are known to cause problems \citep{Skrondal_RabeHesketh_2004, lele_dennis_lutscher_2007, lele_2010, lele_nadeem_schmuland_2010, solymos_2010, lewbel_2019}.
One method that detects non-identified parameters in classical statistical models is data cloning \citep{lele_dennis_lutscher_2007, lele_2010, lele_nadeem_schmuland_2010, solymos_2010}.
Data cloning increases the total sample size by artificially replicating the data and focuses on the standard errors of the parameters.
If the standard error of a parameter does not decrease at a rate of $1/\sqrt{n}$ with increasing sample size $n$, such a parameter is deemed to be non-identified \citep{lele_dennis_lutscher_2007, lele_2010, lele_nadeem_schmuland_2010, solymos_2010, gustafson_15}.

In contrast, non-identification of parameters is not a problem for complex BHMs, because proper priors lead to proper posteriors \citep{gelfand_sahu_99, eberly_carlin_00, gustafson_15}.
However, it is unknown to what extent the posterior estimates of such parameters are determined by the data.
In complex BHMs, it can happen that the posterior is not determined by the data at all but is completely determined by the prior. 
Because the detection of non-identified parameters in BHMs is very challenging, \cite{carlin_louis_1996} and \cite{eberly_carlin_00} suggest that one should avoid BHMs for which the identifiability issues have not been clarified. 
For this reason, a formal method applicable to Bayesian NNHM and BHMs is needed that can quantify the empirical determinacy of the parameters, thus answering the question of to what extent the posterior estimates are determined by the data. 

Recently, \cite{roos_hunanyan_bakka_rue_2020} considered Bayesian NNHM and proposed a two-dimensional method for sensitivity assessment based on the second derivative of the Bhattacharyya coefficient (BC).
In one dimension, they used a factor $f$ to scale the within-study standard deviation provided by the data.
This scaling perturbed the total number of individual patient data provided for Bayesian meta-analysis, which is related to data cloning \citep{lele_nadeem_schmuland_2010, solymos_2010}.
Alternatively, perturbation of the total number of observations could be expressed by a weighted likelihood with $w = 1/f^2$ \citep{bissiri_holmes_walker_2016, holmes_walker_2017}.
An open question remains whether the applicability of the scaling proposed by \cite{roos_hunanyan_bakka_rue_2020} could be generalized to weighting other likelihoods, or extended to general-purpose Bayesian estimation techniques such as the Bayesian numerical approximation INLA \citep{rue_martino_chopin_09}, and Bayesian MCMC sampling with JAGS \citep{plummer_jags_16}.

We expect that likelihood weighting will affect both the location and the spread of the marginal posterior distributions.
Because the method proposed by \cite{roos_hunanyan_bakka_rue_2020} quantifies the total impact of the scaling on the posteriors without focusing on location and spread, a refined method is needed.
Ideally, such a refined method should quantify what proportion of the total impact of the data on the posterior affects the location and what proportion affects the spread.

To provide a refinement, we consider the second derivative of the BC induced by the weighted likelihood, which quantifies the total empirical determinacy (TED) for each parameter.
Based on the properties of both the BC and the rules for the computation of the second derivative, we split TED into two parts: one for location (EDL) and one for the spread (EDS).
This enables the quantification of pEDL, the proportion of EDL to TED, and pEDS, the proportion of EDS to TED.
This method is implemented in the R package \texttt{ed4bhm} and works with JAGS and INLA. 
The method proposed quantifies how much weighted data impact posterior parameter estimates and whether location or spread is more affected by the weighting.

To demonstrate how the proposed method works in applications, we consider two case studies and one simulation study. 
We use INLA and JAGS for the estimation.
We consider both centered and non-centered parametrizations for Bayesian NNHM and demonstrate how TED, pEDL, and pEDS depend on the amount of pooling induced by the heterogeneity prior.
Moreover, we consider both the Bayesian NNHM applied to $\log(\text{OR})$ and a logit model.
To demonstrate that the proposed method is useful in applications, for clarifying the empirical determinacy of parameters, we consider three variants of the NNHM combined with different prior assumptions in a simulation study.

In Section~\ref{sec:methods}, we review the theoretical background behind our method and its implementation through R package \texttt{ed4bhm}. 
In Section~\ref{sec:dataandmodels}, we review the case studies and the models assumed. 
Then, in Section~\ref{sec:results} we present the results which demonstrate the performance of the measures TED, pEDL and pEDS in applications. 
We conclude with a discussion in Section~\ref{sec:Discussion}.

\section{Methods}
\label{sec:methods}
In this section, we review the Bayesian NNHM, present the idea of weighting the likelihood, and define the TED measure, which quantifies the total empirical determinacy of posterior parameters. Moreover, we define the proportion of empirical determinacy for posterior location (pEDL) and spread (pEDS). In addition, we specify the implementation of TED, pEDL and pEDS in INLA and JAGS. 
Finally, a short description of the R package \texttt{ed4bhm} is provided. 

\subsection{Bayesian meta-analysis}
The Bayesian normal-normal hierarchical model (NNHM) consists of three levels: the sampling model, the latent random-effects field, and the priors. 
We consider the Bayesian NNHM with data $y_j, j = 1, \ldots, k$, within-study standard deviations $\sigma_j$ and between-study standard deviation $\tau$
\begin{equation}
\begin{split}
y_{j} \mid \theta_j, \sigma_j & \sim  {\mathcal N}(\theta_{j},  \sigma_j^{2}), \\
\theta_{j}\mid \mu, \tau  &  \stackrel{\text{iid}}{\sim}  {\text N}(\mu, \tau^{2}), \quad j=1,\ldots , k, \\
\mu \sim \pi(\mu), & \quad \tau \sim \pi(\tau).
\end{split}
\label{eq:NNHM}
\end{equation}
In this model, $\sigma_j$ are fixed, $\mu$ is the overall effect, $\tau$ is the heterogeneity, and $\theta_j$ are the random effects.
For case studies, we assume a weakly informative normal prior $\pi(\mu)$ and either a half-normal (HN) or a half-Cauchy (HC) prior $\pi(\tau)$. 
We use HN and HC with domain on $[0, \infty)$ that emerge after taking absolute value of normal and Cauchy distributions located at 0 with domain on $(-\infty, \infty)$.
We estimate marginal posterior distributions for all the parameters $\psi \in \left\{\mu, \tau, \theta_1, \ldots, \theta_k \right\}$ with both INLA \citep{rue_martino_chopin_09} and JAGS \citep{plummer_jags_16}. 

\subsection{Likelihood weighting}
Likelihood weighting is applicable to BHMs with various primary outcomes. 
A base model is the model without weighting ($w=1$). 
Posterior distribution in the base model can be obtained as
\begin{equation}
\pi_1(\eta \mid y) \propto \pi(y \mid \eta, \theta) \pi(\theta \mid \eta) \pi(\eta),
\label{eq:NNHMbase}
\end{equation}
where $\pi(y \mid \eta, \theta)$ is the likelihood, $\pi(\theta \mid \eta)$ is the latent field, $\pi(\eta)$ is the assumed prior distribution, and $\eta$ is the set of all the parameters in the model except the random effects $\theta$.

The posterior distribution in a model with weighted likelihood can be written as
\begin{equation}
\pi_w(\eta \mid y) \propto \left( \pi(y \mid \eta, \theta) \right)^w \pi(\theta \mid \eta) \pi(\eta).
\label{eq:NNHMweighted}
\end{equation}
Note that, $w>1$ gives more weight to the likelihood (corresponds to the case of having more data) and $w<1$ gives less weight to the likelihood (corresponds to the case of having less data).
For example, for the Bayesian NNHM given by \eqref{eq:NNHM}, the posterior distribution from a weighted model reads as
$$\pi_w(\mu, \tau, \theta_1, \ldots, \theta_k \mid y ) \propto (\pi(y \mid \mu, \theta_1, \ldots, \theta_k))^w \pi(\theta_1, \ldots, \theta_k \mid \tau)  \pi(\mu) \pi(\tau),$$ where $\theta_1, \ldots, \theta_k$ are random effects.

\subsection{Affinity measure Bhattacharyya coefficient}
Following \cite{roos_hunanyan_bakka_rue_2020}, we consider the Bhattacharyya coefficient (BC) \citep{bhattacharyya_43} between two probability density functions, which is given by
\begin{equation}
\mathrm{BC}(\pi_1(\psi \mid y), \pi_w(\psi \mid y)) = \int_{-\infty}^{\infty} \sqrt{\pi_1(\psi \mid y) \pi_w(\psi \mid y)} d\psi,
\label{eq:BC}
\end{equation}
where $\psi = (\eta, \theta)$.
BC attains values in [0, 1]. Moreover, BC is 1 if and only if the two densities completely overlap and is 0 when the domains of the two densities have no overlap.
BC has convenient numerical properties and is invariant under any one-to-one transformation (for example, logarithmic) \citep{jeffreys_61, roos_held_11, roos_martins_held_rue_15, roos_hunanyan_bakka_rue_2020}. 
Furthermore, BC is directly connected to the Hellinger distance (H), by $\mathrm{H}^2 = 1 - \mathrm{BC}$.

For two normal densities, BC defined in Equation \eqref{eq:BC} reads
\begin{equation}
\mathrm{BC}(\pi^\text{N}_1, \pi^\text{N}_w) = \sqrt{\frac{2 \sigma_1 \sigma_w}{\sigma^2_1+\sigma^2_w }} \exp \left[ -\frac{ (\mu_w-\mu_1)^2 }{4(\sigma^2_w + \sigma^2_1)} \right],
\label{eq:BCN}
\end{equation}
where $\pi^\text{N}_j$ is the density of a $\mathcal{N}(\mu_j, \sigma^2_j)$ distribution with $j = \{1,w\}$.
In general, any density can be approximated to the first order by a normal distribution \citep{johnson_kotz_balakrishnan_94, hjort_glad_95}. 
Following \cite{roos_hunanyan_bakka_rue_2020}, we approximate the marginal posterior distribution $\pi(\psi \mid y)$ for $\psi \in \{\mu, \tau, \theta_1, \ldots, \theta_k\}$ by $\pi^\text{N}(\psi \mid y)$ and apply moment matching together with the Equation \eqref{eq:BCN} to obtain
\begin{equation}
\mathrm{BC}(\pi_1(\psi \mid y), \pi_w(\psi \mid y)) \approx \mathrm{BC}(\pi^\text{N}_1, \pi^\text{N}_w). 
\end{equation}
For posteriors of precision parameters, we apply this approximation to log-transformed posteriors.
This approach is justified by the invariance of BC.

\subsection{Quantification of the total empirical determinacy}
Following \cite{roos_hunanyan_bakka_rue_2020}, we define the TED($\psi$) measure as the negative second derivative of the Bhattacharyya coefficient (BC) between the base and weighted posterior distributions for each parameter of the model. Here,
\begin{equation}
\text{TED}(\psi) = - \left. \frac{d^2 \text{BC}(\pi_1(\psi \mid y), \pi_w(\psi \mid y) )}{d w^2}  \right|_{w=1},
\label{eq:TED}
\end{equation}
where $\pi_w(\psi \mid y)$ is the marginal posterior distribution from the model with weighted likelihood with $\psi \in \{\mu, \tau, \theta_1, \ldots, \theta_k\}$. 
By $(\text{H}^2)^{\mydprime} = (1 - \text{BC})^{\mydprime} = -\text{BC}^{\mydprime}$, TED evaluates the second derivative of the squared Hellinger distance $\text{H}^2$.
Note that, $w = 1$ denotes the base model with the original likelihood, which induces the base marginal posterior density $\pi_1(\psi \mid y)$. 
We quantify the total empirical determinacy (TED) of the marginal posterior of the parameter.
TED can be used to compare the empirical determinacy between different parameters $\psi$ and $\phi$ by computing a ratio \text{TED}$(\psi)$/\text{TED}$(\phi)$ \citep{roos_hunanyan_bakka_rue_2020}. 

Note that TED in Equation \eqref{eq:TED} measures the impact of likelihood weighting on the whole marginal posterior distribution. 
We refine this approach and consider the impact of likelihood weighting on location (L) and spread (S) of the marginal posterior distribution in the next section.

\subsection{Empirical determinacy of location and spread} 
In this section, we show how to quantify the impact of likelihood weighting on posterior location (L) and spread (S). 
This gives rise to empirical determinacy of location (EDL) and spread (EDS). 

Given the \text{BC} between two normal densities in Equation~\eqref{eq:BCN}, we consider $\mathrm{BC}(w) = \mathrm{BC}_{\text{S}}(w) \mathrm{BC}_{\text{L}}(w) $, where $$\mathrm{BC}_{\text{S}}(w) = \sqrt{\frac{2\sigma_1 \sigma_w}{\sigma_1^2 + \sigma_w^2}}$$ and
$$\mathrm{BC}_{\text{L}}(w) = \exp \left\{- \frac{(\mu_w - \mu_1)^2}{4(\sigma_w^2 + \sigma_1^2)} \right\}.$$
Equations (2)--(5) of the Supplementary Material show that
\begin{equation}
\left.(\mathrm{BC})^{\mydprime}  \right|_{w=1} =  \left.(\mathrm{BC}_{\text{L}})^{\mydprime}  \right|_{w=1} + \left.(\mathrm{BC}_\text{S})^{\mydprime} \right|_{w=1},
\label{eq:property}
\end{equation}
which is equivalent to
\begin{equation}
\text{TED}(\psi) = \text{EDL}(\psi) + \text{EDS}(\psi),
\label{eq:TED_I_plus_B}
\end{equation}
where
\begin{equation}
\text{EDL}(\psi) = - \left. \frac{d^2 \mathrm{BC}_{\text{L}}(\pi_1(\psi \mid y), \pi_w(\psi \mid y) )}{d w^2}  \right|_{w=1},
\label{eq:Bpsi}
\end{equation}
and
\begin{equation}
\text{EDS}(\psi) = - \left. \frac{d^2 \text{BC}_\text{S}(\pi_1(\psi \mid y), \pi_w(\psi \mid y) )}{d w^2}  \right|_{w=1},
\label{eq:Ipsi}
\end{equation}
for $\psi \in \{\mu,\tau,\theta_1, \ldots, \theta_k \}$. 
Equation~\eqref{eq:TED_I_plus_B} facilitates computation of the proportion of the empirical determinacy of the location (pEDL), which is the proportion of EDL to TED, and the empirical determinacy of the spread (pEDS), which is the proportion of the EDS to TED (Equation~\eqref{eq:pH2LStotal})
\begin{equation}
\text{pEDL} (\psi) = \frac{\text{EDL}(\psi)}{\text{TED}(\psi)} \quad \text{and} \quad \text{pEDS} (\psi) = \frac{\text{EDS}(\psi)}{\text{TED}(\psi)},
\label{eq:pH2LStotal}
\end{equation}
where $\psi \in \{\mu,\tau,\theta_1, \ldots, \theta_k \}$.

\subsection{Computational issues}
In practice, we approximate the derivatives numerically by the second-order central difference quotient formula.
Note that $\mathrm{BC}(\pi_1^\text{N}(\psi \mid y), \pi_{1}^\text{N}(\psi \mid y)) = 1$.
Thus, for Equation~\eqref{eq:TED}, we obtain 
\begin{align}
\text{TED}(\psi) & \approx -\frac{d^2 \mathrm{BC}(\pi_1^\text{N}(\psi \mid y), \pi_{w}^\text{N}(\psi \mid y) )}{d w^2}  \\
& \approx \frac{\mathrm{BC}(\pi_1^\text{N}(\psi \mid y), \pi_{1+\delta}^\text{N}(\psi \mid y)) -2  + \mathrm{BC}(\pi_1^\text{N}(\psi \mid y), \pi_{1-\delta}^\text{N}(\psi \mid y))}{\delta^2}. \label{eq:numDiffPlus} 
\end{align}
For computations, we consider weights $w = 1 \pm \delta$, with $\delta = 0.01$.
For precisions, we conduct all computations on the logarithmic scale.

\subsubsection{Implementation of TED in INLA}
In order to implement $\text{TED}(\psi)$ in INLA, we use the INLA object from the base model ($w=1$). 
The function \texttt{inla.rerun} fits an INLA model with weighted likelihood. 
We use this function for $w = 1 - \delta$ and $w = 1+\delta$. 
Moreover, we extract the summary statistics (mean and standard deviation) of marginal posterior distributions directly from INLA objects.
For log-precision, we extract the estimates on internal scale provided in the INLA object.

\subsubsection{Implementation of TED in JAGS}
\label{ssec:fcIS}
For JAGS, we derive a general formula for weighting the likelihood in BHMs fit by MCMC sampling. The posterior distributions for the base and weighted models are given by Equations~\eqref{eq:NNHMbase} and \eqref{eq:NNHMweighted}.
From Equation~\eqref{eq:NNHMbase} we obtain
\begin{equation}
\pi(\eta) \propto \frac{\pi_1(\eta \mid y)}{\pi(y \mid \eta, \theta) \pi(\theta \mid \eta)}.
\label{eq:reformulatedNNHMbase}
\end{equation}
We then plug in the formula \eqref{eq:reformulatedNNHMbase} into \eqref{eq:NNHMweighted} to get
\begin{equation}
\pi_w(\eta \mid y) \propto \left(\pi(y \mid \eta, \theta) \right)^w \pi(\theta \mid \eta) \frac{\pi_1(\eta \mid y)}{\pi(y \mid \eta, \theta) \pi(\theta \mid \eta)}.
\end{equation}
Then, 
\begin{equation}
\pi_w(\eta \mid y) \propto \left(\pi(y \mid \eta, \theta) \right)^{w-1} \pi_1 (\eta \mid y).
\label{eq:fcIS}
\end{equation}

To estimate the posterior from a weighted model without re-running MCMC sampling we use Equation~\eqref{eq:fcIS}. 
To evaluate the likelihood values $\pi(y \mid \eta, \theta)$ in JAGS, we activate the \texttt{dic} module, which monitors
\begin{equation}
\text{dev}(\psi) = -2 \log(\pi(y|\psi)),
\end{equation}
where $\psi = (\eta, \theta)$.
Hence, 
\begin{equation}
\pi(y|\psi) = \exp(-\text{dev}(\psi)/2).
\end{equation}
The MCMC samples $\psi^{(m)}$ from $\pi_1(\psi \mid y)$ and $\text{dev}(\psi)$ values can be extracted from the JAGS object provided by JAGS. 
To estimate the mean and standard deviation of $\pi_w(\psi \mid y)$, we compute
\begin{equation}
\widehat{\E} (\psi \mid y)  = \frac{1}{\sum_{m=1}^M c_m} \sum_{m=1}^M c_m \psi^{(m)}
\label{eq:expIS}
\end{equation}
and
\begin{equation}
\widehat{\SD} (\psi \mid y) = \sqrt{\widehat{\Var}(\psi \mid y) } =\sqrt{ \frac{1}{\sum_{m=1}^M c_m}  \sum_{m=1}^M c_m \left( \psi^{(m)} - \widehat{\E} (\psi \mid y ) \right)^2 },
\label{eq:seIS}
\end{equation}
where $c_m$ = $\left(\pi(y \mid \psi) \right)^{w-1}$ \citep{held_sabanesbove_2020}. 
For log-precisions, we take the logarithmic transformation of the MCMC sample for the precision parameter and apply Equations~\eqref{eq:expIS} and \eqref{eq:seIS} on that transformed sample.

\subsection{Relative latent model complexity}
\label{ssec:rlmc}
In the Bayesian NNHM, the interplay between the within-study standard deviation $\sigma_i$ values provided by the data from $k$ studies and the between-study standard deviation $\tau$ can be characterized by the effective number of parameters in the model \citep{spiegelhalter_best_carlin_linde_02}. Like \cite{roos_hunanyan_bakka_rue_2020}, we consider the standardized ratio, the relative latent model complexity (RLMC)
\begin{equation}
\text{RLMC} = \frac{1}{k}\sum^k_{i = 1} \frac{\tau^2}{\tau^2 + \sigma^2_i}.
\label{eq:rlmc}
\end{equation}
RLMC defined in Equation~\eqref{eq:rlmc} attains values between 0 and 1 and expresses the amount of pooling induced by a heterogeneity prior \citep{gelman_hill_07}. 

In Section~\ref{sec:res8schools}, a grid of scale parameter values (4.1, 10.4, 18, 31.2, 78.4) for the heterogeneity prior $\tau$ induced by the grid of RLMC values fixed at 0.05, 0.25, 0.5, 0.75, 0.95. 
Whereas HN(4.1), which corresponds to RLMC=0.05, assigns much probability mass to $\tau$ values close to 0 and induces much pooling, HN(78.4), which corresponds to RLMC=0.95, assigns less probability mass to values of $\tau$ close to 0 and induces little pooling.

\subsection{Bayesian computation and convergence diagnostics}
The MCMC simulations performed in this paper are based on four parallel chains, with a length of $4 \times 10^5$ iterations. 
In each chain, we removed the first half of iterations as a burn-in period and from the remaining $2 \times 10^5$ iterations we kept every 100th iteration in each of the four chains. 
Our choice of these parameters was guided by \texttt{raftery.diag} for Model C1 (Section 4.4) and supported by \citet{vehtari_gelman_simpson_carpenter_buerkner_2021}.

To assess the convergence to a stationary distribution, we applied Convergence Diagnostics and Output Analysis implemented in the package \texttt{coda} \citep{plummer_best_cowles_vines_96}.  
Moreover, we implemented the rank plots proposed by \citet{vehtari_gelman_simpson_carpenter_buerkner_2021}, which are histograms of posterior draws ranked over all chains and plotted separately for each chain. 
Nearly uniform rank plots for each chain indicate good mixing of chains. 
In addition, we used the function \texttt{n.eff} from the package \texttt{stableGR} \citep{vats_knudson_2021}, which calculates the effective sample size using the lugsail variance estimators and determines whether Markov chains have converged.

\subsection{The R package \texttt{ed4bhm}}
The open-access R package \texttt{ed4bhm} \emph{Empirical determinacy for Bayesian hierarchical models} (\url{https://github.com/hunansona/ed4bhm}) uses the proposed method to quantify the empirical determinacy of BHMs implemented in INLA and in JAGS. The two main functions in this package are called \texttt{ed.inla} and \texttt{ed.jags}. These functions were used to generate the results reported in Sections~\ref{ssec:nm_np} -- \ref{sec:simulation7models}, and in Sections 3 -- 4 and 7 of the Supplementary Material. 

\section{Data and models}
\label{sec:dataandmodels}
In this section, we review data and models for two case studies. Moreover, we review the design of a simulation study. 

\subsection{Eight schools}
\label{ssec:8schools}
To quantify the effect of a coaching program on SAT-V (Scholastic Aptitude Test-Verbal) scores in eight high schools (Table 3 on page 18 of the Supplementary Material), data from a randomized study was pre-analyzed and used for a Bayesian meta-analysis. The data from these eight schools has been used to study the performance of the Bayesian NNHM and to demonstrate typical issues which arise for BHMs \citep{gelman_hill_07, gelman_carlin_stern_dunson_vehtari_rubin_14, vehtari_gelman_simpson_carpenter_buerkner_2021}.

We consider two parametrizations of the Bayesian NNHM: the centered and non-centered parametrizations \citep{vehtari_gelman_simpson_carpenter_buerkner_2021}. The model with centered parametrization is defined as
\begin{equation}
\begin{split}
y_j & \sim \mathcal{N}(\theta_j, \sigma^2_j), \\
\theta_j & \sim \mathcal{N}(\mu, \tau^2),\\
\mu & \sim \mathcal{N}(0, 16), \\
\tau & \sim \text{HN}(5),
\end{split}
\label{8sch:centered}
\end{equation}
where $j = 1, \ldots, 8$.
This parametrization is used for both INLA and JAGS. On the other hand, the model with the non-centered parametrization \citep{gelman_carlin_stern_dunson_vehtari_rubin_14,  vehtari_gelman_simpson_carpenter_buerkner_2021} reads
\begin{equation}
\begin{split}
y_j & \sim \mathcal{N}(\theta_j, \sigma^2_j), \\
\theta_j & = \mu + \tau \tilde{\theta}, \\
\tilde{\theta_j} & \sim \mathcal{N}(0, 1), \\
\mu & \sim \mathcal{N}(0, 16), \\
\tau & \sim \text{HN}(5),
\end{split}
\label{8sch:noncentered}
\end{equation}
for $j = 1, \ldots, 8$, which we implemented in JAGS.

\subsection{Respiratory tract infections}
\label{ssec:rti}
In this section, we review the meta-analysis data set including 22 randomized, controlled clinical trials on the prevention of respiratory tract infections (RTI) by selective decontamination of the digestive tract in intensive care unit patients \citep{bodnar_etal_17} that are presented in Table 4 on page 19 of the Supplementary Material.

For the RTI data set, we consider two different models. First, a Bayesian NNHM model
\begin{equation}
\begin{split}
y_j & \sim \mathcal{N}(\theta_j, \sigma^2_j), \\
\theta_j & = \mu + \eta_j, \quad j = 1, \ldots, 22, \\
\mu & \sim \mathcal{N}(0, 16), \\
\eta_j & \sim \mathcal{N}(0, \tau^{2}), \\
\tau & \sim \text{HC}(1),
\end{split}
\label{rti:NNHM}
\end{equation}
and second, a binomial model with logistic transformation 
\begin{equation}
\begin{split}
z_j & \sim \text{Bin}(p_j, n_j), \\
\text{logit} (p_j) & =  \alpha + \beta x_j  + \eta_j, \quad j = 1, \ldots, 44, \\
\alpha & \sim \mathcal{N}(0, 16), \\
\beta & \sim \mathcal{N}(0, 16), \\
\eta_j & \sim \mathcal{N}(0, \tau^{2}), \\
\tau & \sim \text{HC}(1), \\
\end{split}
\label{rti:extended} 
\end{equation}
where $x$ is a stacked vector attaining value 1 for the treatment and 0 for the control group, $z = (\text{f.t}, \text{f.c})$, $n = (\text{n.t}, \text{n.c})$, where \text{f.t}, \text{f.c}, \text{n.t} and \text{f.c} are vectors of length 22 (see in Table~4 of the Supplementary Material). 
Originally, the weakly informative HC(1) prior assumption for $\tau$ was suggested by \citet{bodnar_etal_17}. \citet{ott_plummer_roos_2020} suggested a prior N(0, 16) for $\mu$ and used the HC(1) prior for $\tau$. We assume priors for $\alpha$ and $\beta$ similar to the prior for $\mu$.

\subsection{Simulation study}
\label{sec:solymos}
Our simulation study extends the original simulation suggested by \citet{solymos_2010}.
We simulate a sample of random observations of size $n = 50$ under NNHM with parameters $\mu = 2.5, \sigma = 0.2, \tau = 0.5$. See Section~7 of the Supplementary Material for further details. 

For the analysis of the simulated data, we use three types of models A, B, and C. Model A is a normal model that does not assume random effects. Models B and C assume a Bayesian NNHM with random effects. 
Model B assumes that the within-study standard deviation is known and assigns a prior to the between-study standard deviation $\tau$. 
This defines the usual Bayesian NNHM. 
In contrast, model C assigns priors to both the within-study ($\sigma$) and between-study ($\tau$) standard deviations. 
Note that the parameters $\sigma$ and $\tau$ in model C are non-identified, because only the sum of the within-study and between-study variances is identified by the likelihood \citep{bayarri_berger_04, lele_2010, solymos_2010, gelman_carlin_stern_dunson_vehtari_rubin_14}.

For the parameter $\mu$, all models A, B, and C assume a zero mean normal prior with variance fixed at $10^3$. For model B, three different priors are assigned to the between-study standard deviation: exp(N(0, $10^3$)) for B1 \citep{ solymos_2010}, SqrtIG(0.001, 0.001) for B2, and SqrtIG(4, 1) for B3. For model C3, a SqrtIG(150, 6) prior is assigned to the within-study standard deviation $\sigma$. Models C1 and C3 assume identical priors for $\tau$ and $\sigma$.
Table~5 on page 22 of the Supplementary Material reports the properties of all priors assumed for the standard deviations. Whereas exp(N(0, $10^3$)) and SqrtIG(0.001, 0.001) show very large medians, SqrtIG(4, 1) and SqrtIG(150, 6) show medians close to both true parameters $\tau=0.5$ and $\sigma=0.2$ chosen for the simulation.

\section{Results}
\label{sec:results}

\subsection{Motivating examples}
\label{ssec:nm_np}
Sections 3 and 4 of the Supplementary Material consider two motivating examples. Both examples use a normal likelihood with identified parameters.
One example considers the posterior of the mean. The other example focuses on the posterior of the precision.
Because the parameters in the likelihoods of both motivating examples are identified, the rate of the decrease of the sample size is close to $1/\sqrt{w}$, which matches the result provided by data cloning \citep{lele_dennis_lutscher_2007, lele_2010, lele_nadeem_schmuland_2010, solymos_2010}. 
In both motivating examples, the estimates of the pEDS are close to 1. 
This indicates that likelihood weighting affects mostly the spread of the parameters. 
In fact, both examples motivate that the $1/\sqrt{w}$ rate used by cloning translates into the properties of the pEDS measure.

\subsection{Eight schools example}
\label{sec:res8schools}

\begin{sidewaystable}[!htpb]
\centering
\begin{tabular}{llr|rrrrr|rrr|rr}
\hline
param & method & ESS & mean & sd &  q0.025  &  q0.5 & q0.975  & TED  & EDL  &  EDS & pEDL  & pEDS \\
\hline
\multirow{3}{*}{$\mu$} & INLA &&3.58 & 2.95 & -2.24 &3.58 & 9.34&  0.11 & 0.09 & 0.02 & 0.82 & 0.18\\
 & JAGSc &1.5e+04 &  3.58 & 2.94 & -2.30 & 3.56 & 9.36&0.10 & 0.08 & 0.02 & 0.82 & 0.18\\
 & JAGSnc &1.8e+04 & 3.59 & 2.96 &  -2.18 & 3.56 & 9.36& 0.11 & 0.09 & 0.02 & 0.83 & 0.17 \\
 &&&&&&&&&&&&\\
 \hline\hline
 &&&&&&&&&&&&\\
 \multirow{3}{*}{$\log(\tau^{-2})$} & INLA &&-1.60 & 2.18 &  -4.49 &-2.07 & 4.00&  7e-04 & 7e-04 & 3e-05 & 0.96 & 0.04\\
 & JAGSc & 1.6e+04& -1.60 & 2.22&  -4.48 & -2.07 & 4.06& 7e-04 & 5e-04  & 2e-04 & 0.69 & 0.31\\
 & JAGSnc &1.6e+04&-1.59 & 2.18 &  -4.44 & -2.07 & 4.06& 6e-04 & 5e-04  & 7e-05 & 0.88 & 0.12\\
\end{tabular}
\caption{\label{tab:medianCrI_inla_jags_8schools_c_and_nc} Model eight schools: mean, sd, 95\%CrI, median and TED, EDL, EDS, pEDL and pEDS estimates for marginal posterior distributions for the parameters $\mu$ and $\log (\tau^{-2})$ calculated in INLA and JAGS (centered and non-centered parametrizations) with ESS of MCMC samples for the data provided in Table 3 on page 18 of the Supplementary Material.}
 \end{sidewaystable}
 
We analyze the data from the eight schools (Table 3 on page 18 of the Supplementary Material) with INLA and JAGS and show the posterior descriptive statistics and estimates of the empirical determinacy for the parameters $\mu$ and $\log (\tau^{-2})$ in Table~\ref{tab:medianCrI_inla_jags_8schools_c_and_nc} on page~\pageref{tab:medianCrI_inla_jags_8schools_c_and_nc}. 
For JAGS, both the centered (JAGSc, Equation~\eqref{8sch:centered}) and non-centered (JAGSnc, Equation~\eqref{8sch:noncentered}) parametrizations showed high values of ESS. 
Although the values of the posterior descriptive statistics provided by INLA, JAGSc, and JAGSnc are similar, it is unknown to what extent the posteriors of $\mu$ and $\log (\tau^{-2})$ are impacted by the data. 

This issue is addressed by the estimates of both the total empirical determinacy (TED) and the within-parameter empirical determinacy (EDL and EDS). For example, the values of TED($\mu$) are larger than those of TED($\log(\tau^{-2})$). 
This indicates that the data impacts the posterior of $\mu$ more than the posterior of $\log(\tau^{-2})$. pEDL and pEDS provide further details and demonstrate that for both $\mu$ and $\log(\tau^{-2})$ the location of the marginal posterior distribution is more impacted by the data than its spread. Whereas the estimates of pEDS($\mu$) provided by INLA, JAGSc and JAGSnc are close to 0.18, those of pEDS($\log(\tau^{-2})$) differ depending on the centered (JAGSc and INLA) and non-centered (JAGSnc) parametrization. This result indicates that the parametrization used for MCMC sampling may affect the impact of the data on the parameter estimates.

\begin{figure}
\begin{knitrout}
\definecolor{shadecolor}{rgb}{0.969, 0.969, 0.969}\color{fgcolor}
\includegraphics[width=1\linewidth,height=0.8\textheight]{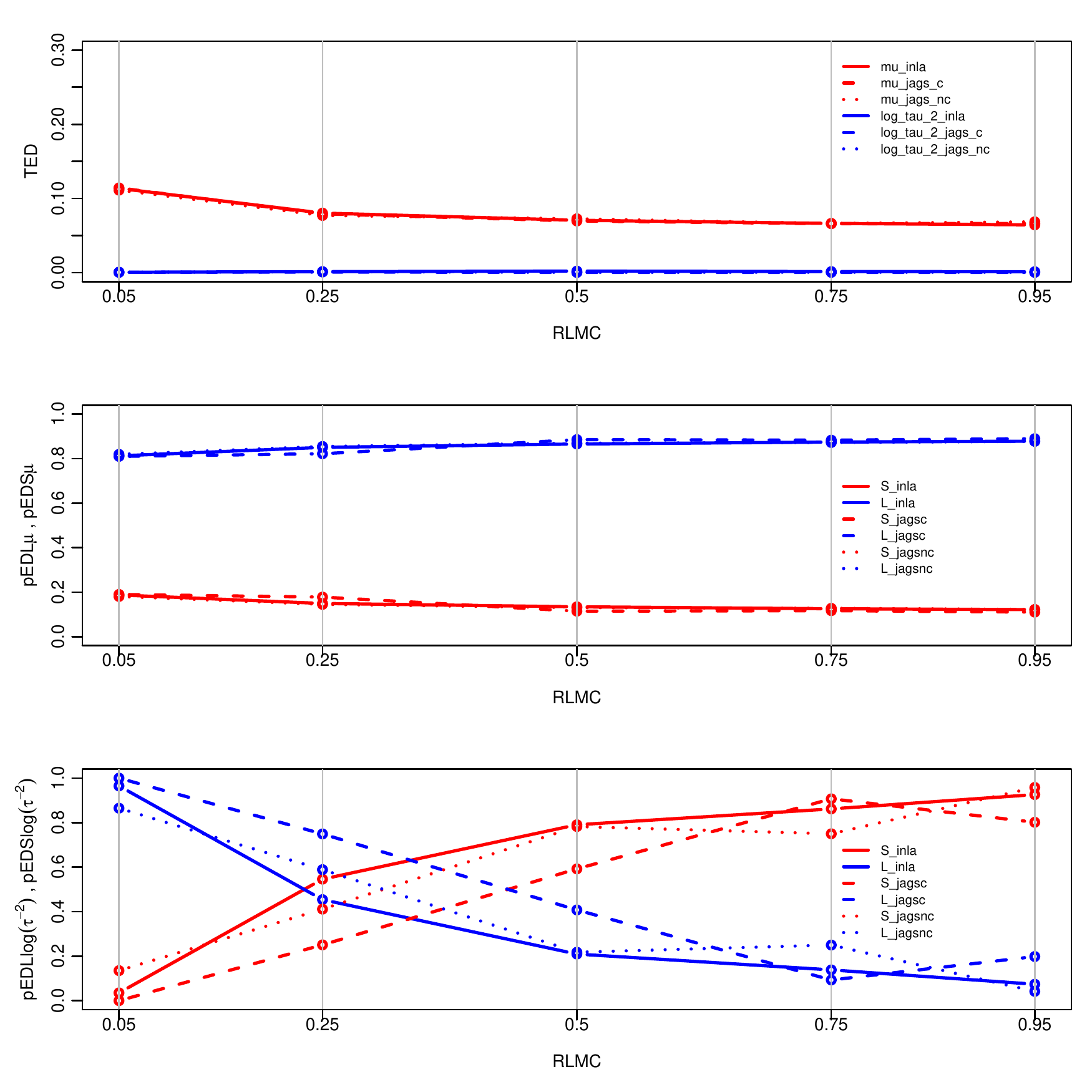} 

\end{knitrout}
\caption{\label{fig:transph8sch} Model eight schools: Transition phase plots for the total TED measure (top) and for the pEDL and pEDS of $\mu$ (middle) and of $\log (\tau^{-2})$ (bottom) for the data in Table 3 on page 18 of the Supplementary Material modeled according to Equations~ \eqref{8sch:centered} and \eqref{8sch:noncentered} (centered and non-centered parametrizations) and analyzed with INLA and JAGS. The effective median RLMC for eight schools data with HN(5) heterogeneity prior is 0.08. The scale parameters for the HN prior across the grid of 0.05, 0.25, 0.5, 0.75, 0.95 RLMC values are equal to 4.1, 10.4, 18, 31.2, 78.4 (Section~\ref{ssec:rlmc}). }
\end{figure}

Figure~\ref{fig:transph8sch} puts the results of Table~\ref{tab:medianCrI_inla_jags_8schools_c_and_nc} on page~\pageref{tab:medianCrI_inla_jags_8schools_c_and_nc} in a broader context and shows a transition phase plot for the data of the eight schools (Table 3 on page 18 of the Supplementary Material) fit by INLA, JAGSc, and JAGSnc across a grid of HN heterogeneity priors with scale parameters equal to 4.1, 10.4, 18, 31.2, 78.4. 
As explained in Section~\ref{ssec:rlmc}, this grid specifies an RLMC grid (0.05, 0.25, 0.5, 0.75, 0.95) of pooling induced by the heterogeneity prior.
The top panel of Figure~\ref{fig:transph8sch} demonstrates that TED is always larger for $\mu$ (red) than for $\log(\tau^{-2})$ (blue) across both the grid of RLMC values and the estimation techniques (INLA, JAGSc, and JAGSnc). 
Independently of the amount of pooling induced by the heterogeneity prior, the posterior of $\mu$ is more determined by the data than is the posterior of $\log(\tau^{-2})$. 

The proportion of the estimates of the within-parameter empirical determinacy (pEDS) for the scale of $\mu$ (middle panel) and $\log(\tau^{-2})$ (bottom panel) of Figure~\ref{fig:transph8sch} provide more insight. 
For HN heterogeneity priors with values of the scale parameter equal to 4.1, 10.4, 18, 31.2, 78.4 and for three estimation techniques (INLA, JAGSc, JAGSnc), the values of pEDS($\mu$) remain at a low level of at most 20\%. 
This indicates that the data determine the spread of the posterior of $\mu$ less than its location.
In contrast, the estimates of pEDS($\log(\tau^{-2})$) highly depend on the amount of pooling induced by the heterogeneity prior. 
The estimates of pEDS($\log(\tau^{-2})$) increase from 10\% to 90\% across the values of RLMC. 
This means that heterogeneity priors that induce little pooling lead to posteriors of $\log(\tau^{-2})$ with spreads more determined by the data than for heterogeneity priors that induce much pooling. 
Although the estimates of pEDS($\mu$) are similar for the three estimation techniques (INLA, JAGSc, JAGSnc), the values of pEDS($\log(\tau^{-2})$) depend on the estimation technique. 

\subsection{RTI example}
\label{sec:resrti}

\begin{sidewaystable}[!htpb]
\centering
\begin{tabular}{llrr|rrrrr|rrr|rr}
\hline
param & method & model & ESS & mean & sd & q0.025 & q0.5 & q0.975 & TED & EDL & EDS & pEDL & pEDS \\
\hline
&&&&&&&&&&&&&\\
\multirow{2}{*}{$\mu$} & \multirow{5}{*}{NNHM} & INLA && -1.29 & 0.21 & -1.74 &-1.28 & -0.91&  0.39 & 0.39 & 7e-04 & 1.00 & 2e-03\\
 && JAGS & 1.6e+04 &  -1.29 & 0.21 & -1.74 & -1.28 & -0.91& 0.40 & 0.40 & 6e-04 & 1.00 & 2e-03\\
 &&&&&&&&&&&&&\\
\cline{5-14}
 &&&&&&&&&&&&&\\
 \multirow{2}{*}{$\log(\tau^{-2})$} && INLA & &0.79 & 0.57 &  -0.27 &0.76 & 2.00 &  1.04 & 0.78 & 0.27 & 0.74 & 0.26\\
 && JAGS & 1.6e+04 &0.79 & 0.59&  -0.27 & 0.76 & 2.04 & 1.23 & 0.78  & 0.45 & 0.63 & 0.37\\
 &&&&&&&&&&&&\\
 \hline
 \hline
 &&&&&&&&&&&&\\
 \multirow{2}{*}{$\alpha$} & \multirow{5}{*}{logit} & INLA &  &-0.59 & 0.26 & -1.11 &-0.60 & -0.07 &  0.02 & 0.02 & 0.01 & 0.75 & 0.25\\
 && JAGS & 1e+04  &  -0.60 & 0.26 & -1.11 & -0.60 & -0.08 & 0.02 & 0.01 & 6e-03 & 0.61 & 0.39\\
&&&&&&&&&&&&&\\
\cline{5-14}
&&&&&&&&&&&&&\\
 \multirow{2}{*}{$\beta$} & \multirow{5}{*}{logit} & INLA &  &-1.50 & 0.38 & -2.25 &-1.49 & -0.76 &  0.03 & 0.03 & 3e-03 & 0.90 & 0.10\\
 && JAGS & 1e+04  &  -1.49 & 0.38 & -2.26 & -1.48 & -0.74& 0.02 & 0.02 & 3e-03 & 0.89 & 0.11\\
 &&&&&&&&&&&&&\\
\cline{5-14}
 &&&&&&&&&&&&&\\
 \multirow{2}{*}{$\log(\tau^{-2})$} && INLA & & -0.30 & 0.27 &  -0.85 &-0.30 & 0.21&  0.33 & 0.32 & 6e-03 & 0.98 & 0.02\\
 && JAGS & 1.6e+04  &-0.31 & 0.27&  -0.86 & -0.30 & 0.21 & 0.31 & 0.30  & 0.01 & 0.96 & 0.04\\
&&&&&&&&&&&&&\\
\end{tabular}
\caption{ \label{tab:medianCrI_inla_jags_rti_NNHM} Models for RTI: NNHM (above) and logit (below). Mean, sd,  95\% CrI, median and TED, EDL, EDS, pEDL and pEDS for marginal posterior distributions for the parameters $\mu$ and $\log (\tau^{-2})$ for NNHM, and $\alpha$, $\beta$ and $\log (\tau^{-2})$ for logit models defined in Equations \eqref{rti:NNHM} and \eqref{rti:extended} calculated in INLA and JAGS for data provided in Table~4 on page 19 of the Supplementary Material with ESS of MCMC samples.}
\end{sidewaystable}

Table~\ref{tab:medianCrI_inla_jags_rti_NNHM} on page~\pageref{tab:medianCrI_inla_jags_rti_NNHM} provides posterior descriptive statistics and estimates of the empirical determinacy for the RTI data (Table~4 of the Supplementary Material) fit by INLA and JAGS. 
We considered two different models: a Bayesian NNHM defined in Equation~\eqref{rti:NNHM} with a normal primary outcome and a Bayesian logit model defined in Equation~\eqref{rti:extended} with a binomial primary outcome.
The MCMC chains provided by JAGS show high values of ESS (Table~\ref{tab:medianCrI_inla_jags_rti_NNHM} on page~\pageref{tab:medianCrI_inla_jags_rti_NNHM}).
The marginal posteriors provided by INLA and JAGS match well and provide similar descriptive statistics (see Figures 6 and 7 of the Supplementary Material).
The descriptive statistics for the parameter $\log(\tau^{-2})$ differ between the NNHM and logit models.
For both the NNHM and logit models, the estimates of TED indicate that the posterior of $\log(\tau^{-2})$ is more impacted by the data than are the posteriors of $\mu$, $\alpha$, and $\beta$.
NNHM provides lower values of pEDS($\mu$) than the values of pEDS($\beta$) provided by the logit model.
This indicates that the structure of the primary outcome and the model can have a great impact on the empirical determinacy of the parameters.

\subsection{Simulation: empirical determinacy}
\label{sec:simulation7models}

\begin{sidewaystable}[htpb!]
\centering
\begin{tabular}{l|cccc|cccccccc}
  \hline
 & $\mu_\text{TED}$ & $\sigma^{*}_\text{TED}$ & $\tau^{*}_\text{TED}$ & $\gamma^{*}_\text{TED}$ & $\mu_\text{pEDL}$ & $\mu_\text{pEDS}$ & $\sigma^{*}_\text{pEDL}$ & $\sigma^{*}_\text{pEDS}$ & $\tau^{*}_\text{pEDL}$ & $\tau^{*}_\text{pEDS}$ & $\gamma^{*}_\text{pEDL}$ & $\gamma^{*}_\text{pEDS}$ \\ 
  \hline
$\text{A}:\text{INLA}$ & 0.15 &  &  & 0.16 & 6e-08 & 1.00 &  &  &  &  & 0.14 & 0.86 \\ 
  $\text{A}:\text{JAGS}$ & 0.15 &  &  & 0.15 & 5e-04 & 1.00 &  &  &  &  & 0.09 & 0.91 \\ 
  $\text{B}1:\text{INLA}$ & 1e-06 &  & 0.21 &  & 0.00 & 1.00 &  &  & 0.89 & 0.11 &  &  \\ 
  $\text{B}1:\text{JAGS}$ & 7e-04 &  & 0.21 &  & 0.35 & 0.65 &  &  & 0.92 & 0.08 &  &  \\ 
  $\text{B}2:\text{INLA}$ & 8e-06 &  & 0.24 &  & 0.00 & 1.00 &  &  & 0.90 & 0.10 &  &  \\ 
  $\text{B}2:\text{JAGS}$ & 3e-05 &  & 0.27 &  & 0.05 & 0.95 &  &  & 0.89 & 0.11 &  &  \\ 
  $\text{B}3:\text{INLA}$ & 2e-04 &  & 0.13 &  & 6e-08 & 1.00 &  &  & 0.93 & 0.07 &  &  \\ 
  $\text{B}3:\text{JAGS}$ & 3e-03 &  & 0.12 &  & 0.05 & 0.95 &  &  & 0.95 & 0.05 &  &  \\ 
  $\text{C}1:\text{INLA}$ & 0.45 & 3e+03 & 2e+03 &  & 6e-08 & 1.00 & 0.90 & 0.12 & 0.88 & 0.13 &  &  \\ 
  $\text{C}1:\text{JAGS}$ & 182.48 & 1e+04 & 9e+03 &  & 0.70 & 0.25 & 0.57 & 0.65 & 0.2 & 0.92 &  &  \\ 
  $\text{C}2:\text{INLA}$ & 0.01 & 1e+04 & 1e+04 &  & 6e-08 & 1.00 & 0.83 & 0.37 & 0.94 & 0.12 &  &  \\ 
  $\text{C}2:\text{JAGS}$ & 0.06 & 6e+02 & 2e+02 &  & 0.8 & 0.2 & 0.67 & 0.34 & 0.87 & 0.13 &  &  \\ 
  $\text{C}3:\text{INLA}$ & 3e-04 & 1.02 & 0.18 &  & 6e-08 & 1.00 & 1.00 & 0.00 & 0.92 & 0.08 &  &  \\ 
  $\text{C}3:\text{JAGS}$ & 4e-03 & 1.13 & 0.2 &  & 1e-03 & 1.00 & 0.99 & 0.01 & 0.97 & 0.03 &  &  \\ 
   \hline
\end{tabular}
\caption{\label{table:TED} TED, pEDL, and pEDS values from models A, B.1, B.2, B.3, C.1, C.2 and C.3 for the simulated data described in Section \ref{sec:solymos}. $\sigma^{*}$, $\tau^{*}$ and $\gamma^{*}$ stand for $\log(\sigma^{-2})$, $\log(\tau^{-2})$ and $\log(\gamma^{-2})$, respectively.} 
\end{sidewaystable}

\begin{table}[ht]
\centering
\begin{tabular}{rrrrr}
  \hline
 & $\mu$ & $\sigma$ & $\tau$ & $\gamma$ \\ 
  \hline
A & 16000.00 &  &  & 16000.00 \\ 
  B1 & 16000.00 &  & 16000.00 &  \\ 
  B2 & 16000.00 &  & 16000.00 &  \\ 
  B3 & 16000.00 &  & 16000.00 &  \\ 
  C1 & 78.65 & 2203.84 & 2826.11 &  \\ 
  C2 & 15429.11 & 9314.01 & 9079.20 &  \\ 
  C3 & 16000.00 & 16000.00 & 15678.21 &  \\ 
   \hline
\end{tabular}
\caption{\label{table:ESS} ESS for the JAGS samples for the models A, B.1, B.2, B.3, C.1, C.2 and C.3 for the simulation study described in Section \ref{sec:solymos}. For the MCMC sampling the total number of iterations used is $4 \times 10^5$, the burn-in is $2 \times 10^5$, thinning = 100.} 
\end{table}

Table~\ref{table:TED} on page~\pageref{table:TED} provides estimates of the empirical determinacy for the simulation study described in Section~\ref{sec:solymos}, which considers three types of NNHM models (A, B and C), which are fit by INLA and JAGS. 
The results for JAGS are based on MCMC samples with ESS reported in Table~\ref{table:ESS} on page ~\pageref{table:ESS}.

For model A, the estimates of TED($\mu$) and TED($\log(\gamma^{-2})$) in Table~\ref{table:TED} on page~\pageref{table:TED} are similar. 
Moreover, pEDS($\mu$) and pEDS($\log(\gamma^{-2})$) are high. 
For example, pEDS$(\log(\gamma^{-2}))\ge 0.86$ demonstrates that the spread of the posterior of $\log(\gamma^{-2})$ is highly impacted by the data. 

For models B1, B2 and B3, the values of TED($\log(\tau^{-2})$) are larger than the values of TED($\mu$) in Table~\ref{table:TED} on page~\pageref{table:TED} and indicate that the posterior of $\log(\tau^{-2})$ is more impacted by the data than is the posterior of $\mu$. 
Similarly to the phase transition in the example of the eight schools discussed in Section~\ref{sec:res8schools}, the values of TED($ \log(\tau^{-2})$) depend on the heterogeneity prior. 
Moreover, the large estimates of pEDS($\mu$) demonstrate that a large proportion of the change in the posterior distribution of $\mu$ is due to the change in spread rather than in location. 
In contrast, the low estimates of pEDS($\log(\tau^{-2})$) indicate that the posterior spread is not much determined by the data.

Models of type C assume priors on the between-study standard deviation and on the within-study standard deviation.
The estimation of models C1, C2 and C3 is very challenging. 
For example, the ESS of MCMC samples for model C1 provided by JAGS (Table~\ref{table:ESS} on page~\pageref{table:ESS}) are very small.
Although the parameters $\sigma$ and $\tau$ are non-identified in models C2 and C3, ESS is reasonably high.
Table~\ref{table:TED} on page~\pageref{table:TED} shows that TED of the posteriors of $\log(\sigma^{-2})$  and $\log(\tau^{-2})$ is larger than that of the parameter $\mu$. 
Again, the values of TED($ \log(\sigma^{-2})$)  and TED($ \log(\tau^{-2})$) depend on heterogeneity priors. 
For C1 and C2, the estimates of pEDS($\mu$) differ between INLA and JAGS attaining large values for INLA and low values for JAGS. 
This indicates that numerical Bayesian approximation (INLA) and MCMC sampling (JAGS) can react differently to models with a non-identified likelihood. 
For model C1, pEDS($\log(\sigma^{-2})$) and pEDS($\log(\tau^{-2})$) values shown by INLA and JAGS disagree. 
This is due to the small values of the ESS value of the MCMC samples provided by JAGS. In contrast, these  estimates of pEDS agree well for models C2 and C3. 
The difference between the values of pEDS($\log(\sigma^{-2})$) and pEDS($\log(\tau^{-2})$) for models C2 and C3 indicates that the prior assumptions impact the values of pEDS.

\section{Discussion}
\label{sec:Discussion}
We considered two case studies and one simulation study. 
For the well-known eight-school example we applied Bayesian NNHM, considering INLA analysis and both centered and non-centered parametrizations for JAGS. 
Moreover, we provided a transition phase plot for the estimates of TED, pEDL and pEDS across a grid of heterogeneity prior scale parameters, which govern the amount of pooling induced by the heterogeneity prior. 
This provided insights into how TED, pEDL and pEDS change depending on the properties of the heterogeneity prior. 
For the RTI data set, we used both the Bayesian NNHM applied to log(OR) and a logit model providing insights into how TED, pEDL and pEDS change depending on the primary outcome. 
To challenge the method proposed, we relaxed the assumption that the within-study standard deviation is known and assumed priors for both the within-study and between-study standard deviation. 
This provided an insight into how TED, pEDL and pEDS perform when the underlying model has two non-identified parameters for both INLA and JAGS. 
The proposed method provided novel insight and proved useful in clarifying the empirical determinacy of the parameters in the Bayesian NNHM.

The analysis of two simple motivating examples, normal mean and normal precision, translated the results provided by data cloning, which are based on the $1/\sqrt{n}$ criterion, to the unified pEDS measure. 
They showed that for identified parameters of non-hierarchical likelihoods the spread of the posterior is mainly affected by the likelihood weighting and leads to pEDS estimates close to 1. 
We prefer the use of pEDS, because the application of the $1/\sqrt{n}$ criterion to BHMs is not well justified \citep{jiang_2017, lewbel_2019}. 

This method considerably refines and generalizes the original method proposed by \cite{roos_hunanyan_bakka_rue_2020}.
We proposed a unified method that quantifies what proportion of the total impact of likelihood weighting on the posterior is due to the change in the location (pEDL) and what proportion is due to the change in the spread (pEDS).
This was achieved by matching posteriors moments with those of a normal distribution for the computation of the BC.
Note that the normal distribution encapsulates the information about its location and spread in two unrelated parameters.
This property proved useful for the definition of pEDL and pEDS.
Because likelihood weighting affects both the location and the spread of the marginal posterior distributions, the proposed method is better suited for the quantification of empirical determinacy than other methods that are focused only on the total impact \citep{roos_hunanyan_bakka_rue_2020, kallioinen_paananen_buerkner_vehtari_2021}.

We successfully applied the proposed method to a non-normal likelihood. 
This shows that the proposed method can also be applied to other BHMs with different primary outcomes. 
Moreover, we implemented this method in general-purpose Bayesian estimation programs, such as INLA and JAGS. 
For JAGS, we developed and applied a technique for the fast and efficient computation of likelihood weighting, which dispenses with re-runs of MCMC chains.  
All these refinements and generalizations enable future extensions of the method proposed to complex BHMs and to other general-purpose Bayesian estimation programs, such as Stan \citep{Stan_development_team_Manual_16}.
The estimates of the empirical determinacy will help researchers understand to what extent posterior estimates are determined by the data in complex BHMs.

Currently, the weighting approach focuses on weights that are very close to 1. 
This is very useful to study the impact of the total number of patients, which is induced by within-study sample size changes, on the posteriors of parameters in Bayesian NNHM. 
However, the data cloning approach, which changes the total number of studies included for the Bayesian meta-analysis, indicates that integer weights that are larger than 1 could also be very useful in assessing the impact of the data on BHMs. 
More work is necessary to extend the proposed method to cope with integer weights.

The proposed method is based on estimates of the mean and standard deviation for location and spread.
These posterior descriptive statistics are provided by default by general-purpose software for Bayesian computation.
Moreover, they are used by other modern approaches to quantify the impact of priors on posteriors \citep{reimherr_meng_nicolae_2021}.
However, for stability reasons, \cite{vehtari_gelman_simpson_carpenter_buerkner_2021} recommend the use of location and spread estimates based on quantiles.
In addition, \cite{vehtari_gelman_gabry_2017} recommend Pareto smoothed importance sampling (PSIS) to regularize importance weights, which flow into the computation of the location and spread of the weighted posteriors.
Therefore, anchoring the method proposed on quantiles and the use of PSIS are two additional goals that need to be addressed by our future research.

This paper focuses mainly on Bayesian NNHM. 
However, the application of the method proposed to other complex BHMs would provide a further insight into the empirical determinacy of posterior parameter estimates in other applications. 
The openly accessible R package \texttt{ed4bhm} (\url{https://github.com/hunansona/ed4bhm}) conveniently facilitates the application of the proposed method in other settings.

\section*{Acknowledgments}
Support by the Swiss National Science Foundation (no. 175933) granted to Ma{\l}gorzata Roos is gratefully acknowledged.

\bibliographystyle{plainnat}
\bibliography{ed}
\end{document}